\documentstyle[preprint,prd,aps]{revtex}

\def\frc#1#2{{\textstyle{#1 \over #2}}}
\def\beq{\begin{equation}}
\def\eeq{\end{equation}}

\begin{document}

\small
\draft

%%%%%%%%%%%%%%%%%%%%%%%%%%%%%%%%%%%%%%%%%%%%%%%%%%%%%%%%%%%%%%%%%%%%%%
%% Title Page %%%%%%%%%%%%%%%%%%%%%%%%%%%%%%%%%%%%%%%%%%%%%%%%%%%%%%%%
%%%%%%%%%%%%%%%%%%%%%%%%%%%%%%%%%%%%%%%%%%%%%%%%%%%%%%%%%%%%%%%%%%%%%%

\preprint{MPI-PhT/96-42}

\title{SN 1987A Gamma-Ray Limits on the Conversion of 
Pseudoscalars}

\author{Jack W.~Brockway and Eric D.~Carlson}
\address{Olin Physical Laboratory, Wake Forest University\\ 
Winston-Salem, NC 27109, USA}

\author{Georg G.~Raffelt}
\address{Max-Planck Institut f\"{u}r Physik 
(Werner-Heisenberg-Institut)\\
F\"ohringer Ring 6, 80805 M\"unchen, Germany}

\date{\today}

\maketitle

\bigskip\bigskip

\centerline{\bf Abstract}

Pseudoscalar particles $\phi$ usually couple electromagnetically by an
interaction of the form $\frc14 g \phi F {\widetilde F}$, allowing
them to convert to photons in the presence of magnetic fields.
Notably, new low-mass pseudoscalars emitted from supernova (SN) 1987A
would have been converted to $\gamma$-rays in the intervening magnetic
field of the galaxy. Therefore, measurements by the Solar Maximum
Mission (SMM) Gamma-Ray Spectrometer (GRS) can limit the inverse
coupling constant to $g^{-1}>1{\times}10^{11}\,\rm GeV$, assuming the
pseudoscalar is massless.  This is an improvement over other astrophysical
limits of a factor of about 2.5.

\vfill\eject

%%%%%%%%%%%%%%%%%%%%%%%%%%%%%%%%%%%%%%%%%%%%%%%%%%%%%%%%%%%%%%%%%%%%%%
%% Section I  %%%%%%%%%%%%%%%%%%%%%%%%%%%%%%%%%%%%%%%%%%%%%%%%%%%%%%%%
%%%%%%%%%%%%%%%%%%%%%%%%%%%%%%%%%%%%%%%%%%%%%%%%%%%%%%%%%%%%%%%%%%%%%%

\section{Introduction}

Light neutral pseudoscalars arise naturally as a result of
spontaneously broken global symmetries \cite{kim}. They will be truly
massless if the symmetry is not anomalous. Such particles generically
couple to the electromagnetic field through an interaction of the form
\begin{equation}
        {\cal L}= \frc14 g \phi F^{\mu\nu}{\widetilde F_{\mu\nu}} \; ,
        \label{ffdual}
\end{equation}
where $\widetilde F_{\mu\nu}$ is the dual of the electromagnetic
field-strength tensor $F^{\mu\nu}$ and $g$ is a coupling constant of
dimension (energy)$^{-1}$ which is related to the scale of symmetry
breaking. This coupling leads to the interconversion of photons and
pseudoscalars in external electric or magnetic fields.  Just as the
pion decay constant was originally measured by this Primakoff effect,
it is natural to investigate the existence of new pseudoscalars by
this method. For example, two beautiful experiments to search for the
presence of dark-matter axions in our galaxy are currently under
way~\cite{axiondetection}.

If new low-mass bosons exist, stars will be powerful sources for their
production \cite{axionreviews,book}. Therefore, a classic way to
constrain the coupling $g$ is from observational limits on anomalous
stellar energy losses. The observed helium-burning lifetime of
horizontal-branch stars yields a limit
$g^{-1}\agt1.7{\times}10^{10}\,\rm GeV$, applicable if the particle
mass is not much larger than the stellar core temperature of about
$10\,\rm keV$ and thus covers the important case of axions
\cite{book,hbstars}.

However, if the pseudoscalar mass is very small, more restrictive
limits can be obtained
from the possible conversion of stellar particle fluxes
in the large-scale magnetic field of the galaxy.  This effect would
lead to apparent x-ray or $\gamma$-ray fluxes from stars since the
emitted pseudoscalars have energies representative of the interior
temperature. Essentially, then, one would observe photons from the
stellar core which were converted there to pseudoscalars in the
electric fields of the charged particles of the heat bath, and which
were then back-converted into photons by the galactic magnetic
field. One of us (E.C.)  has previously applied this argument to the
red supergiant $\alpha$-Ori (Betelgeuse) and found a limit
$g^{-1}>4{\times}10^{10}\,\rm GeV$ \cite{carlson}. Later, one of us
(G.R.) pointed out that this method could also be applied to the
pseudoscalar flux from supernova (SN) 1987A \cite{book}.  The
Gamma-Ray Spectrometer (GRS) on the Solar Maximum Mission (SMM)
satellite has set very restrictive limits on a possible $\gamma$-ray
burst in conjunction with the observed SN~1987A neutrino signal, a
fact which has been used to derive limits on neutrino radiative decays
\cite{neutrinodecays}. We presently show that these observations yield
a constraint $g^{-1}>1{\times}10^{11}\,\rm GeV$, more restrictive
than all previous bounds. Of course, it will apply only to massless
pseudoscalars (true Goldstone bosons) and thus excludes axions for
which the horizontal-branch star limit remains the most restrictive.

In Section~2 we discuss the theory of pseudoscalar production in the
supernova's interior.  In Section~3 we discuss their conversion to
photons in the presence of the galaxy's magnetic field and the limit
which such conversion can achieve.  Observational limits on converted
photons by the Solar Maximum Mission satellite is discussed in 
Section~4.  We summarize our findings in Section~5.

%%%%%%%%%%%%%%%%%%%%%%%%%%%%%%%%%%%%%%%%%%%%%%%%%%%%%%%%%%%%%%%%%%%%%%
%% Section II %%%%%%%%%%%%%%%%%%%%%%%%%%%%%%%%%%%%%%%%%%%%%%%%%%%%%%%%
%%%%%%%%%%%%%%%%%%%%%%%%%%%%%%%%%%%%%%%%%%%%%%%%%%%%%%%%%%%%%%%%%%%%%%

\section{Supernova Production of Pseudoscalars}

Pseudoscalars with a coupling of the form (\ref{ffdual}) are produced
primarily by the Primakoff process. In vacuum, its cross section is
logarithmically infinite due to the infinite range of the Coulomb
potential. In a medium, however, Debye screening cuts off this
divergence and leads to a conversion rate per unit time for photons to
pseudoscalars of \cite{primakoff}
\begin{equation}
      \Gamma = {g^2 \kappa^2 T \over 32 \pi} 
         \left[ \left( 1 + {\kappa^2 \over 4E^2} \right)
         \ln \left(1 + {4E^2 \over \kappa^2} \right) -1 \right] \; ,
         \label{primak}
\end{equation}
where $E$ is the photon energy, $T$ the temperature, and $\kappa$ the
inverse Debye screening length.  The overall factor of $\kappa^2$ in
this equation is simply a convenient way of writing the density of
scattering targets, and has nothing especially to do with Debye
screening.  The other appearances of $\kappa^2$ arise from the finite
range of the electric field surrounding charged particles in the
plasma.

In order to derive an expression for $\kappa$ which is appropriate for
the conditions of a SN core, we note that the relevant charged
particles are equal number densities of electrons and
protons. However, because of their large mass difference the electrons
are very degenerate, while the protons are nearly nondegenerate, or at
most partially degenerate. Therefore, electrons are essentially
unavailable as scattering targets because their phase space is almost
completely Pauli blocked. Further, their degeneracy causes them to
form a ``stiff'' background which is difficult to polarize so that
they can be ignored for screening as well.  Because of the large
temperature, the plasma coupling parameter for the protons is much
smaller than unity in spite of the large density so that the plasma is
weakly coupled in the electromagnetic sense.  Therefore, $\kappa$ is
given by the Debye formula
\begin{equation}
       \kappa^2 = {4 \pi \alpha \over T} n_p
       \; , \label{Debye}
\end{equation}
where $n_p$ is the number density of the protons and $\alpha$ is the
fine-structure constant.

Multiplying Eq.~(\ref{primak}) with the density of thermal photons, we
find the pseudoscalar volume production rate $\dot n_\phi$ per unit
energy to be
\begin{equation}
{d \dot n_\phi \over dE} = {g^2\xi^2 T^3 \over 8\pi^3(e^{E/T}-1)}
\left[(E^2 + \xi^2 T^2)\ln (1 + E^2/\xi^2 T^2) - E^2\right] \; ,
\end{equation}
where $\xi^2 \equiv \kappa^2/4T^2$. We stress that it is possible
to treat photons essentially as massless particles because the plasma
frequency for the conditions at hand is small compared with the
temperature.

In order to calculate the total expected flux we use the numerical
SN model S2BH\_0 of Keil, Janka, and Raffelt \cite{model} which
is a representative case. We had numerical details at hand for $t=1$,
5, and 10 seconds after core bounce.  The values of $T$ and $\xi^2$ as
a function of radius are plotted in Fig.\ 1.  Further, in Fig.\ 2 we show
$T/E_{\rm F}$, where the nonrelativistic proton Fermi energy is given
by $E_{\rm F} =p_{\rm F}^2/2m_N$. (The Fermi momentum is related to the
proton density by $n_p=p_{\rm F}^3/3\pi^2$.)  Degeneracy effects
typically become truly important only for $E_{\rm F}/T\agt3$ so that
our approximation of treating protons as nondegenerate is quite
reasonable.

We then integrate over the volume of the star to get the total rate
$d\dot N_\phi/dE$ at which particles are produced.  It is graphed in
Fig.~3 for a coupling strength of $g^{-1} = 10^{10}\,$GeV.  For the
earliest time, $t=1\,$sec, we may be slightly overestimating the
production because of the proton degeneracy.

Once the pseudoscalars are produced, they will escape the supernova
provided their mean free path $\lambda$ for backconversion exceeds
the size of the SN core.  The backconversion rate is just twice
the rate given in Eq.~(\ref{primak}).  Typical photon and pseudoscalar
energies will be around $E=3T$, so that typically $E/2\kappa = 3/\xi$.
In this range the square bracket in Eq.~(\ref{primak}) is of order
unity. Therefore, $\lambda^{-1}\approx g^2\kappa^2 T/16\pi$, almost
independently of the energy.  If we use $T=30\,$MeV and
$\kappa=50\,$MeV as characteristic values, we find $\lambda\approx
g_{10}^{-2}\,10^{12}\,\rm cm$ where $g_{10}^{-1}\equiv
g^{-1}/10^{10}\,{\rm GeV}$.  Therefore, in the range of coupling
constants which is of interest here, the pseudoscalars escape freely
from the SN core once produced.

%%%%%%%%%%%%%%%%%%%%%%%%%%%%%%%%%%%%%%%%%%%%%%%%%%%%%%%%%%%%%%%%%%%%%%
%% Section III %%%%%%%%%%%%%%%%%%%%%%%%%%%%%%%%%%%%%%%%%%%%%%%%%%%%%%%
%%%%%%%%%%%%%%%%%%%%%%%%%%%%%%%%%%%%%%%%%%%%%%%%%%%%%%%%%%%%%%%%%%%%%%

\section{Conversion in the Galactic Magnetic Field}

Having produced our pseudoscalars, we now wish to detect them.  To do
so, we will back-convert them in the galactic magnetic field. In a
field which is roughly homogeneous on scales much larger than the
pseudoscalar wavelength, this conversion process can be viewed as an
oscillation phenomenon much like that of neutrino oscillations
\cite{RaffeltStodolsky}. However, oscillation phenomena will manifest
themselves only if the momentum transfer $(m_\phi^2-m_\gamma^2)/2E$
which is needed to convert a pseudoscalar into a photon exceeds the
inverse scale over which the $B$-field is roughly homogeneous. For
photons, the ``mass'' is given by the plasma frequency $m_\gamma^2=
4\pi\alpha n_e/m_e$ where $n_e$ of order $0.03\,{\rm cm}^{-3}$ is the
interstellar electron density. It corresponds to
$m_\gamma=0.6{\times}10^{-11}\,\rm eV$ or, with a photon energy
$E=100\,\rm MeV$, to $m_\gamma^2/2E=2{\times}10^{-31}\,\rm eV
=(30\,{\rm Mpc})^{-1}$. The relevant spatial extent of the magnetic
field will turn out to be of order $1\,\rm kpc$ so that photons can be
viewed as massless for the present purposes in spite of the plasma
effect. Similarly, for the pseudoscalars to count as effectively
massless we need to require $(m_\phi^2/2E)^{-1}\agt 1\,\rm kpc$ or
$m_\phi\alt 10^{-9}\,\rm eV$. 

If photons and pseudoscalars are effectively massless in this sense,
the conversion probability is given by
\beq
P = \frc14 g^2 B_\perp^2 \ell^2 \; ,
\eeq
where $B_\perp$ is the magnetic field perpendicular to the line of
sight, and $\ell$ is the distance over which the magnetic field is
effectively constant.

The magnetic field of the galaxy has considerable
structure.  There is certainly a toroidal component with a magnitude
of about $2\,\mu$G and a coherence length of about $1\,\rm kpc$.
There are also other components, which may be roughly characterized as
a random contribution of magnitude $5\,\mu$G with a coherence length
of perhaps 10 pc.  Because the random component cannot contribute
coherently, we anticipate that the toroidal field will be the dominant
contribution.  Hence we will assume that there is a constant field of
$2\,\mu$G.  Since $1\,\rm kpc$ is also the approximate thickness of
the disk where we expect this galactic magnetic field, we will assume
a constant field on this scale, which cuts off suddenly after a
distance of $1\,\rm kpc$.  
There may also be extragalactic fields between us
and the Large Magellanic Cloud; if so, they will only increase the
effect.  Obviously, there is considerable uncertainty in this
calculation, and hence our final result can not be taken as precise.

SN~1987A is at a galactic latitude $b=-32.1^\circ$ and longitude
$l=279.6^\circ$.  This unfortunately means that we are looking
primarily along the direction of the magnetic field, so that
\beq
P = \frc14 g^2 B^2 \ell^2(1-\cos^2b\sin^2l) = 
\frc14\, 0.30\, g^2B^2\ell^2 \; .
\eeq
The photon flux at Earth is then given by
\beq
{d\Phi \over dE} = {d\dot N_\phi \over dE} 
{g^2 B_\perp^2 \ell^2 \over 4\pi D^2} \; ,
\eeq
where $\ell = 1\,$kpc is the length of the conversion region,
$D=50\,$kpc is the distance to SN~1987A, and
$B_\perp^2=0.30\,(2\,\mu$G$)^2$.  For a coupling $g^{-1}=10^{10}$GeV,
this yields the fluxes plotted in Fig.~3.

%%%%%%%%%%%%%%%%%%%%%%%%%%%%%%%%%%%%%%%%%%%%%%%%%%%%%%%%%%%%%%%%%%%%%%
%% Section IV %%%%%%%%%%%%%%%%%%%%%%%%%%%%%%%%%%%%%%%%%%%%%%%%%%%%%%%%
%%%%%%%%%%%%%%%%%%%%%%%%%%%%%%%%%%%%%%%%%%%%%%%%%%%%%%%%%%%%%%%%%%%%%%

\section{Detection of Gamma-Rays}

Not surprisingly, no detector was oriented specifically in the
direction of SN 1987A at the time of core collapse.  Fortunately, the
Gamma Ray Spectrometer (GRS) on the Solar Maximum Mission (SMM)
satellite, though pointed in the direction of the Sun, was still able
to detect gamma rays coming from the direction of SN 1987A \cite{GRS}.
The precise timing of the core collapse is known from the observation
of neutrinos by the Irvine-Michigan-Brookhaven \cite{IMB}\ and
Kamiokande II \cite{KII}\ detectors.  Because electron neutrinos and
photons are both effectively massless they should arrive
simultaneously at Earth.  Thus we know exactly which period to search
for an excess of photons from SN~1987A.

During the 10.24 seconds that the neutrinos were detected from SN
1987A, the GRS searched for photons in three energy bins, 4.1--6.4
MeV, 10--25 MeV, and 25--100 MeV.  The 95\%\ confidence limits on the
total fluence of photons during this time period were 0.9, 0.4, and
0.6 $\gamma\,{\rm cm}^{-2}$ respectively \cite{GRS}. Because of the
rising shape of our spectrum, the best limit for us comes from
considering the highest of these bins.  The detailed limit depends on
the shape of the spectrum, and the limit cited in \cite{GRS}\ assumes
a spectrum which falls as $E^{-2}$.  Without knowing more about the
energy dependence of the detector's response, we cannot get a precise
limit, but we will assume that the limit 0.6 $\gamma\,{\rm cm}^{-2}$
applies.

In the highest energy bin, the $\gamma$-ray flux from SN~1987A can be
found by integrating the rates plotted in Fig.~3.  For a coupling of
$g^{-1}=10^{10}$GeV, this turns out to be 2880, 2120, and $1180\,\gamma
\,\rm cm^{-2}s^{-1}$, at $t=1$, 5, and 10 seconds, respectively.  Recall
that at $t=1$ second, there should be a noticeable reduction of the
rate from that calculated due to Pauli blocking of the final-state
protons.  We expect this blocking to be even larger at earlier times,
so we ignore all contributions from $t<1\,$second, and assume that our
calculations are accurate over the rest of the time period.  A linear
fit gives a fluence (time-integrated flux) of $1.8{\times}10^4\,\gamma\,
\rm cm^{-2}$ for the period from $t=1$ to $10$ seconds.

Nominally, this implies a limit of $g^{-1}>1.32{\times}10^{11}\,$GeV.
However, there are considerable uncertainties involved in the
calculation, principally in the details of the galactic magnetic
field, and the energy dependence of the detector response.  If we
assume that there is an uncertainty of a factor of three in our
calculated fluence, then our limit is weakened to 
$g^{-1}>1.0{\times}10^{11}\,$GeV.  Note that since the photon flux
is proportional to $g^4$, the $g$ limit is rather insensitive to small
errors in the flux or detection calculation.

%%%%%%%%%%%%%%%%%%%%%%%%%%%%%%%%%%%%%%%%%%%%%%%%%%%%%%%%%%%%%%%%%%%%%%
%% Section V %%%%%%%%%%%%%%%%%%%%%%%%%%%%%%%%%%%%%%%%%%%%%%%%%%%%%%%%%
%%%%%%%%%%%%%%%%%%%%%%%%%%%%%%%%%%%%%%%%%%%%%%%%%%%%%%%%%%%%%%%%%%%%%%

\section{Discussion and Summary}

We have found that the absence of excess $\gamma$-rays from SN~1987A
limits the photon coupling of a nearly massless pseudoscalar
$g^{-1}>1{\times}{10^{11}}\,{\rm GeV}$. This limit is significantly
more restrictive than $g^{-1}>1.7{\times}{10^{10}}\,{\rm GeV}$ which
is found from the classic energy-loss argument applied to
horizontal-branch stars \cite{book,hbstars}. However, the latter
argument is valid for pseudoscalar masses of up to about $10\,\rm keV$
and thus covers the important case of axions. Of course, for axions
even more restrictive limits obtain from their coupling to nucleons on
the basis of the energy-loss argument applied to SN~1987A. Our present
result is primarily useful for hypothetical massless Goldstone bosons
which do not exhibit significant couplings to ``normal'' fermions such
as electrons or nucleons. If they had such couplings one would
typically expect that these couplings yield more restrictive limits on
the underlying scale of symmetry breaking. Ignoring such couplings
also justifies that we focused on the Primakoff effect to estimate
the mean free path in the SN core. In principle, of course, arbitrary
pseudoscalars could be trapped by interactions with nucleons or
electrons.

In a previous paper \cite{carlson}, one of us (E.C.)  considered what
limit had been set or could be set by looking for $\gamma$-rays from
stars, particularly the nearby supergiant $\alpha$-Ori (Betelgeuse).
A limit of $g^{-1} > 4 {\times}10^{10}\,\rm GeV$ was set by
reexamining old data, and it was pointed out that this limit could be
improved to the level of $1{\times}10^{11}\,\rm GeV$ if a dedicated
observation were made. We have achieved this goal with the existing
SN~1987A data.

%%%%%%%%%%%%%%%%%%%%%%%%%%%%%%%%%%%%%%%%%%%%%%%%%%%%%%%%%%%%%%%%%%%%%%
%% Acknowledgments %%%%%%%%%%%%%%%%%%%%%%%%%%%%%%%%%%%%%%%%%%%%%%%%%%%
%%%%%%%%%%%%%%%%%%%%%%%%%%%%%%%%%%%%%%%%%%%%%%%%%%%%%%%%%%%%%%%%%%%%%%

\acknowledgements 

We would like to thank S.~Harder for his helpful input into this
project.  E.C.\ was supported, in part, by a Sloan Foundation Fellowship.
In Munich, this research was supported, in part, by grant SFB 375 of the
Deutsche Forschungsgemeinschaft. 

%%%%%%%%%%%%%%%%%%%%%%%%%%%%%%%%%%%%%%%%%%%%%%%%%%%%%%%%%%%%%%%%%%%%%%
%% References  %%%%%%%%%%%%%%%%%%%%%%%%%%%%%%%%%%%%%%%%%%%%%%%%%%%%%%%
%%%%%%%%%%%%%%%%%%%%%%%%%%%%%%%%%%%%%%%%%%%%%%%%%%%%%%%%%%%%%%%%%%%%%%

\def\apj#1#2#3{, {\it Ap.\ J.} {\bf #1}, #2 (19#3)}
\def\prd#1#2#3{, {\it Phys.\ Rev.\ D} {\bf #1}, #2 (19#3)}
\def\plb#1#2#3{, {\it Phys.\ Lett.\ B} {\bf #1}, #2 (19#3)}
\def\prl#1#2#3{, {\it Phys.\ Rev.\ Lett.} {\bf #1}, #2 (19#3)}

%%%%%%%%%%%%%%%%%%%%%%%%%%%%%%%%%%%%%%%%%%%%%%%%%%%%%%%%%%%%%%%%%%%%%%
%% Figures %%%%%%%%%%%%%%%%%%%%%%%%%%%%%%%%%%%%%%%%%%%%%%%%%%%%%%%%%%%
%%%%%%%%%%%%%%%%%%%%%%%%%%%%%%%%%%%%%%%%%%%%%%%%%%%%%%%%%%%%%%%%%%%%%%

%\bigskip\bigskip

\newpage

\centerline{{\bf Figure Captions}}

\def\figg#1{\bigskip\bigskip\hang\noindent {\bf Fig.\ #1.}}

\figg1 The temperature $T$ (1a) and the parameter $\xi$ (1b) as
functions of radius for times $t$=1, 5, and 10 seconds, based on the
model S2BH\_0 of Keil, Janka and Raffelt\cite{model}.

\figg2 The parameter $T/E_{\rm F}$ (proton Fermi energy $E_{\rm F}$)
as a function of radius $r$ for $t$=1, 5, and 10 seconds after core
bounce, based on the model S2BH\_0 of Keil, Janka and
Raffelt\cite{model}.

\figg3 The number of pseudoscalars produced (left scale) and resulting
$\gamma$-ray photon flux at the Earth (right scale) for a coupling
$g^{-1}=10^{10}$GeV for times $t$=1, 5, and 10 seconds.

\vfill\eject

\begin{references}

\bibitem{kim}{For a review of light pseudoscalars and their couplings,
see, for example, J.E. Kim, {\it Phys.\ Rep.} {\bf 150}, 1 (1987).}

\bibitem{axiondetection}{C.~Hagmann {\it et al.}, in: {Dark Matter in
Cosmology, Clocks, and Tests of Fundamental Laws (Proceedings of the
XXXth Rencontre de Moriond, Villars-sur-Ollon, Switzerland, January
22--29, 1995)}, edited by B.~Guiderdoni, G.~Greene, D.~Hinds, and
J.~Tr\^an Thanh V\^an (Editions Fronti\'eres, Gif-sur-Yvette, 1995)
pg.~181; S.~Matsuki and I.~Ogawa, {\it ibid.}\ pg.~187.}

\bibitem{axionreviews}{For reviews on axions in astrophysics see
M.S. Turner, {\it Phys. Rep.} {\bf 197}, 67 (1990); G.G. Raffelt, 
{\it Phys. Rep.} {\bf 198}, 1 (1990).}

\bibitem{book}{G.G.~Raffelt, {\it Stars as Laboratories for
Fundamental Physics} (University of Chicago Press, 1996).}

\bibitem{hbstars}{G.G.\ Raffelt and D.S.P.\
Dearborn\prd{36}{2211}{87}.}

\bibitem{carlson}{E.D. Carlson\plb{344}{245}{95}.}

\bibitem{neutrinodecays}{E.W.~Kolb and M.S.~Turner\prl{62}{509}{89};
L.~Oberauer {\it et al.}, {\it Astropart.\ Phys.} {\bf 1}, 377 (1993).}

\bibitem{primakoff}{G.G. Raffelt\prd{33}{897}{86}.}
%See also G.G. Raffelt, {\it Phys. Rep.} {\bf198}, 1 (1990).

\bibitem{model}{W. Keil, H.-T. Janka and G.G.
Raffelt\prd{51}{6635}{95}.}

\bibitem{RaffeltStodolsky}{G.G.~Raffelt and
L.~Stodolsky\prd{37}{1237}{88}.}

\bibitem{GRS}{E.L.~Chupp, W.T.~Vestrand and
C.~Reppin\prl{62}{505}{89}.}

\bibitem{IMB}{K.~Hirata {et al.}\prl{58}{1490}{87}.}

\bibitem{KII}{R.M.~Bionta {\it et al.}\prl{58}{1494}{87}.}

\end{references}
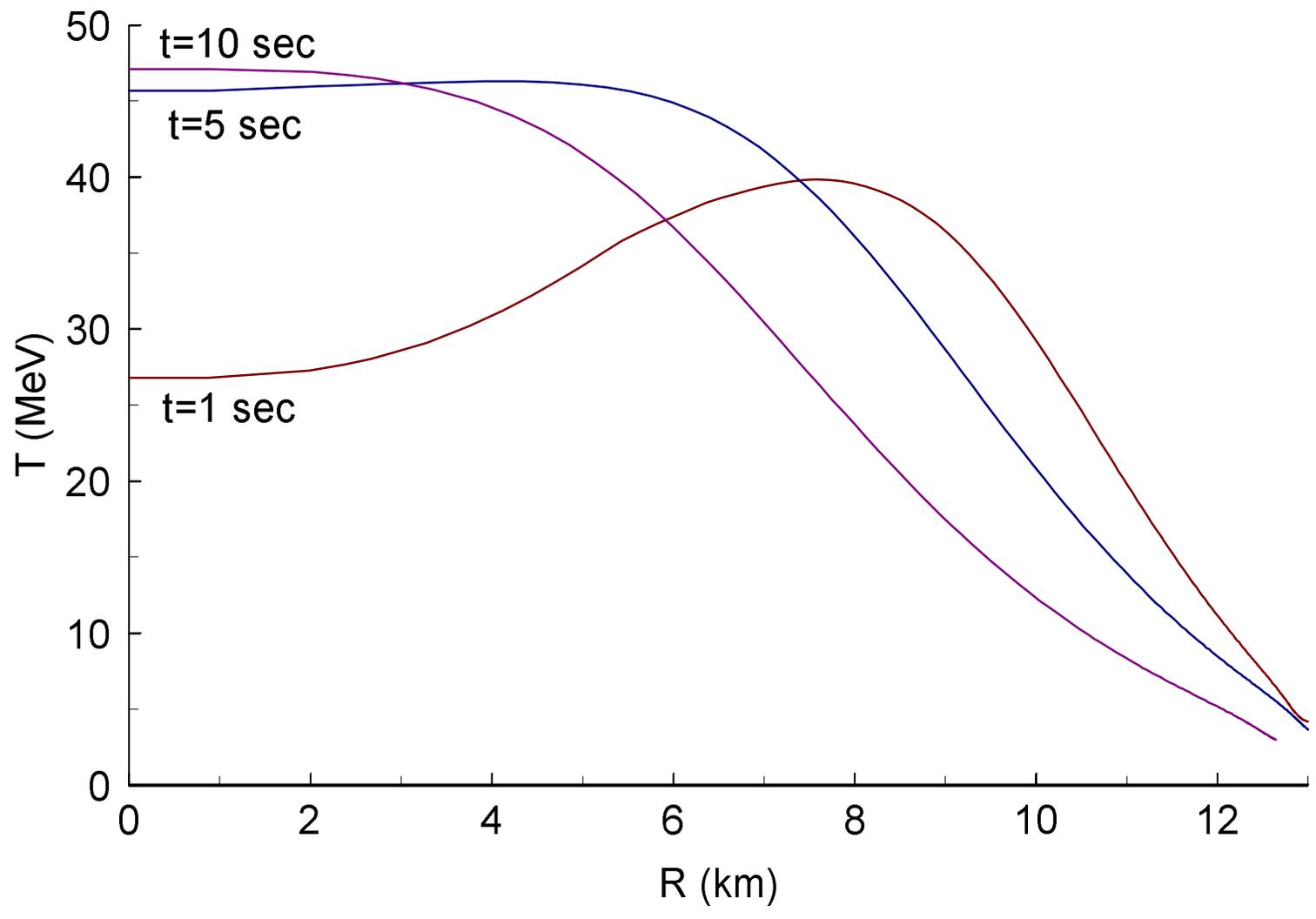
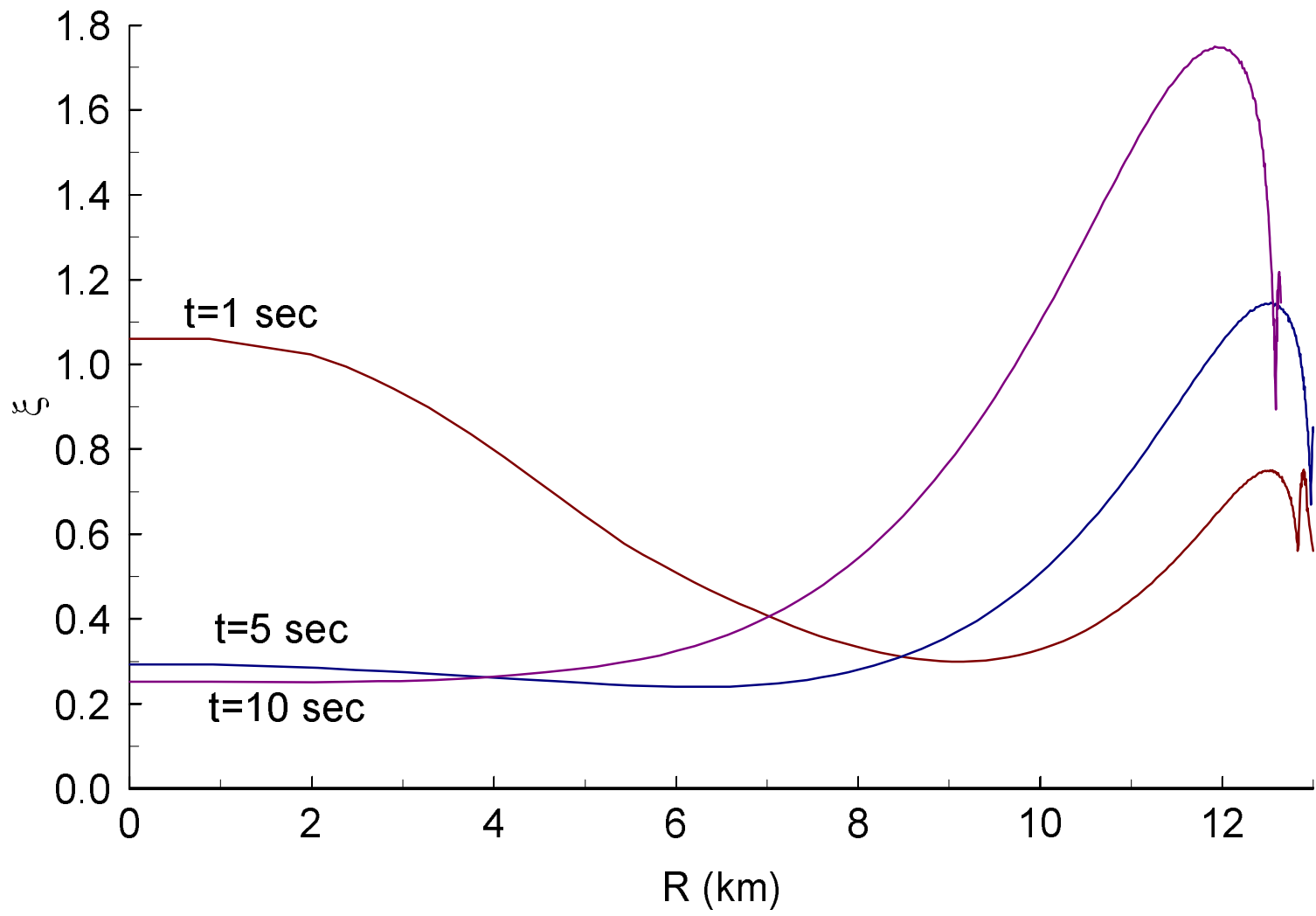
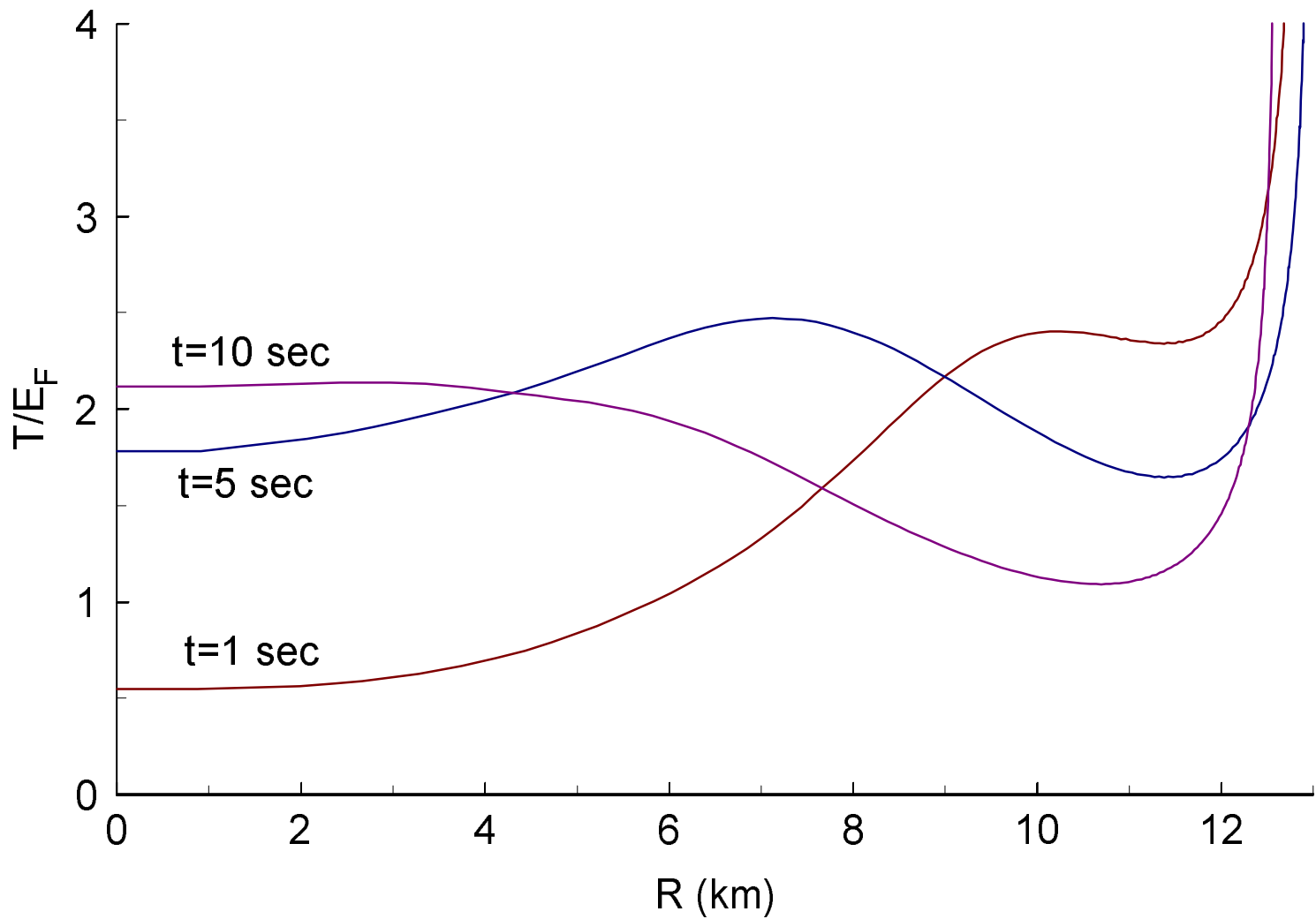
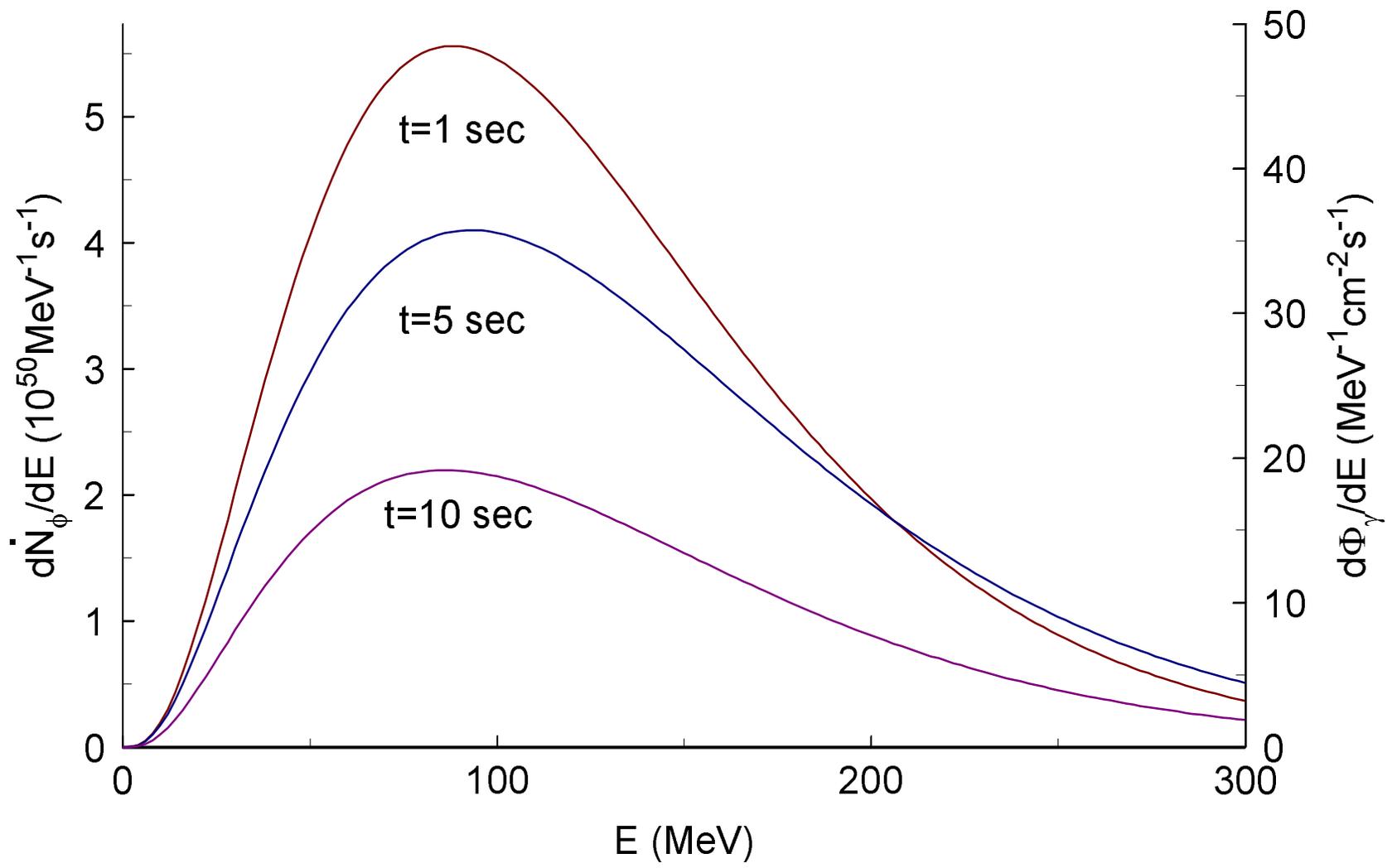
\end{document}